\documentstyle[12pt]{article}
\makeatletter
\def\section{\@startsection {section}{1}{\z@}{3.5ex plus 1ex minus 
 .2ex}{2.3ex plus .2ex}{\large\bf}}
\def\lsim{\mathrel{\mathpalette\gl@align<}}
\def\gsim{\mathrel{\mathpalette\gl@align>}}
\def\gl@align#1#2{\lowe.6ex\vbox{\baselineskip\z@skip\lineskip\z@
    \ialign{$\m@th#1\hfil##\hfil$\crcr#2\crcr\sim\crcr}}}
\makeatother 
\newcommand{\newsection}[1]{
\setcounter{equation}{0}
\section{#1}
}
\newcommand{\be}{\begin{equation}}
\newcommand{\ee}{\end{equation}}
\newcommand{\bea}{\begin{eqnarray}}
\newcommand{\eea}{\end{eqnarray}}
\newcommand{\vs}[1]{\vspace{#1 cm}}

\def\p{\partial}
\def\f{\frac}

\def\a{\alpha}

\def\bl{\biggl}
\def\br{\biggr}

\def\f#1#2{\frac{#1}{#2}}
\def\non{\nonumber}

\begin{document}
\begin{titlepage}
\vs{1}
\begin{center}
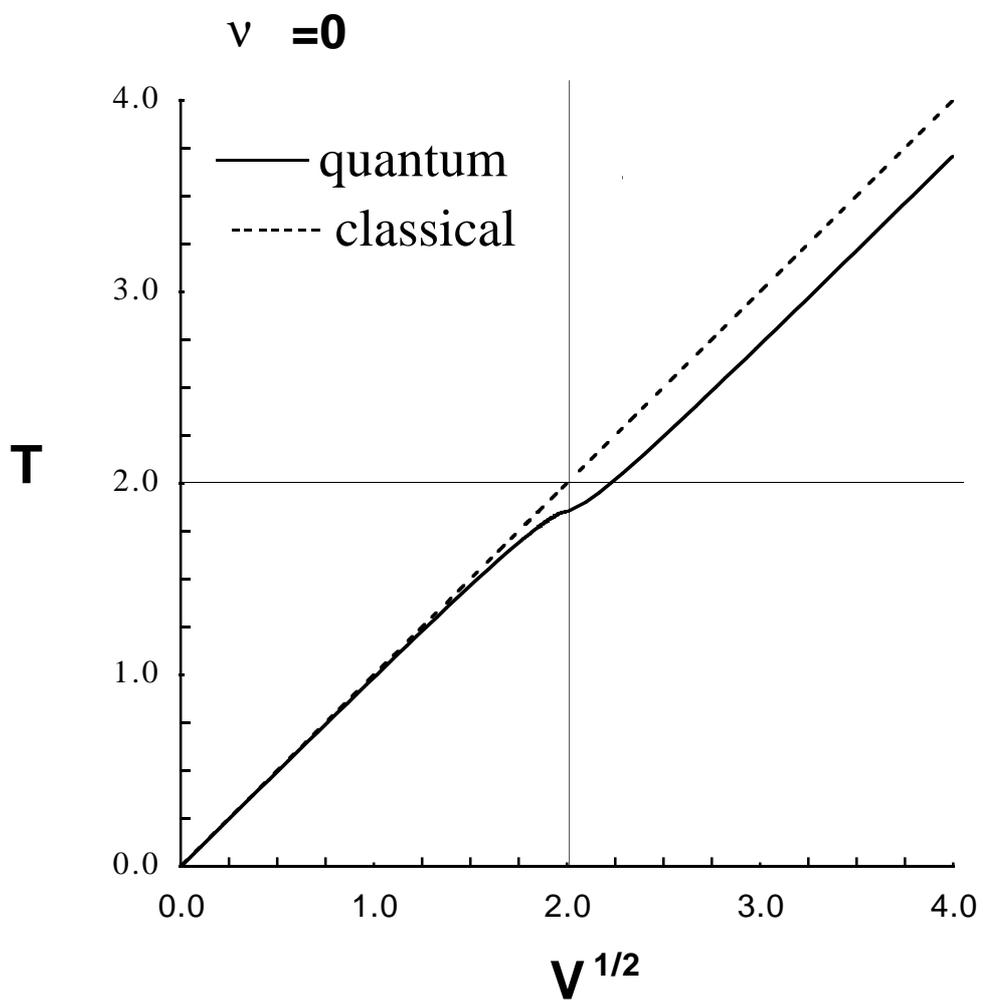
{\Large{\bf de Broglie-Bohm Interpretation for the Wave Function 
             of Quantum Black Holes 
       }
} 
\end{center}
\vspace{1cm}
\setlength{\footnotesep}{15pt}
\begin{center}
{\sc Masakatsu Kenmoku \footnote{e-mail address:kenmoku@cc.nara-wu.ac.jp} 
, Hiroto Kubotani \footnote{e-mail address:kubotani@aquarius.nara-edu.ac.jp} 
, Eiichi Takasugi \footnote{e-mail address:takasugi@phys.wani.osaka-u.ac.jp} 
 \\ and  Yuki Yamazaki \footnote{e-mail address:yamazaki@phys.nara-wu.ac.jp} 
}\break
\\
{\it  $\ ^1$ Department of Physics, Nara Women's University,~Nara 630,~Japan \\ 
 $\ ^2$ Department of Mathematics, Nara University of Education,~Nara 630,~Japan \\ 
 $\ ^3$ Department of Physics, Osaka University,~Osaka 560,~Japan \\
 $\ ^4$ Graduate School of Human Culture, Nara Wemen's University,~Nara 630,~Japan  
}
\end{center}
\vs{1}
\begin{abstract}
We study the quantum theory of the spherically symmetric black holes.
The theory yields the wave function inside the apparent horizon, where the role of  
time and space coordinates is interchanged. 
The de Broglie-Bohm interpretation is applied to the wave function and then  
the trajectory picture on the minisuperspace is introduced in the quantum as 
well as the semi-classical region. 
Around the horizon large quantum fluctuations on the trajectories of metrics 
$U$ and $V$ appear in our model, 
where the metrics are functions of time variable $T$ and are expressed as  
$\displaystyle  ds^2=-\f{{\alpha}^2}{U}dT^2+UdR^2 +Vd\Omega^2$.
On the trajectories, 
the classical relation $U=-V^{1/2}+2Gm$ holds, 
and the event horizon $U=0$ corresponds to the classical apparent horizon 
on $V=2Gm$.  
In order to investigate the quantum fluctuation near the horizon,  
we study a null ray on the dBB trajectory  
and compare it with the one in the classical black hole geometry.
\end{abstract}
\vspace{0.3cm}
\hspace{0.8cm} PACS number(s): 03.65.Bz, 04.60.Ds, 04.70.Dy, 97.60.Lf
\end{titlepage}
\newpage
%
\newsection{Introduction}
%
%
A quantum state with fluctuations of spacetime is described by a state 
vector $\Psi$, which is a solution to the Wheeler-DeWitt (WD) equation 
$H\Psi=0$, where $H$ is the Hamiltonian of the gravitating system\cite{Wheeler,DeWitt67}.
This equation implies that the wave function is a stationary state with zero energy, 
and there is no Schr$\ddot{{\rm o}}$dinger evolution of the physical state, that is,
we cannot get the dynamical picture from this equation. 
This is called the problem of time in quantum gravity 
and has been researched extensively\cite{Isham92}. 
In addition to this, 
there are other problems on the interpretation for the wave function to be 
considered\cite{Kuchar93}. 
In particular, problems occur when we apply the ordinary 
Copenhagen interpretation to the wave function of the whole universe. 
First one is that the observer is also an element of the quantum mechanical system. 
Second one is that in the Copenhagen interpretation, 
one must consider many times of measurements 
performed on a pure ensemble, each element of which is characterized by 
the same state 
$\mid \Psi >$, and when a measurement of observable 
A is made by an external observer described by the classical mechanics, 
the measurement causes the discontinuous change brought about 
by the collapse of the wave function into an eigenstate 
$\mid a >$ with eigenvalue a. It is assumed that the quantity 
$\mid <a \mid \Psi > \mid^2$     
gives the probability for obtaining the measured value a. In quantum 
cosmology, however, since the wave function given by the WD equation 
describes a unique and whole system, one cannot accept the concept of the probability.

Many people use the WKB approach to the quantum cosmology to 
overcome the problem of time\cite{Banks85,Halliwell87}. 
In the standard WKB approach, the system is assumed to 
separated into two parts: 
the gravitational part as the semi-classical system and 
the matter part as the quantum system. 
The time is introduced through the identification of  
the gradient of the classical action with the velocity variable. 
The Hamiltonian of quantum matter  plays the time developing operator  
 and the Schr$\ddot{{\rm o}}$dinger-like equation holds, 
though the total system does not develop the time. 
Some ambiguities and difficulties of this method were pointed out 
\cite{Singh89,Alwis94,Kenmoku97}. 
Though the WKB approach successfully introduce the time and 
the Schr$\ddot{{\rm o}}$dinger equation, the problem of observation remains unsolved 
because the standard WKB approach gives no alternative to the stand point of the Copenhagen 
interpretation. 

The de Broglie-Bohm (dBB) interpretation\cite{deBroglie,Bohm52} 
introduces the time in the very similar 
way to the WKB approach whereas it takes very different interpretation 
to the wave function.  
The dBB interpretation defines a kind of trajectory by identifying the 
momentum with the gradient of the phase of the wave function.
We shall call it the dBB or quantum trajectory. This interpretation seems 
to be able to avoid some of the problems mentioned above 
\cite{Bell87,Holland93}. 
For the problem of the dynamical evolution, we can obtain it through  
the dBB trajectories, which describe the quantum evolution using the time 
coordinate appeared in the original Lagrangian. 
These trajectories are assumed to be the real entity.  
The quantum effects on the trajectories are represented 
by the quantum potential quantitatively
which is defined by the second derivative of the amplitude of the wave function. 
By this quantum potential the dBB trajectories are modified from the classical ones.   
If the quantum potential is small enough compared with the ordinary potential, 
the classical system is realized whether the observer exists or not. 
Therefore the dBB interpretation needs no observer and no collapse 
of the wave function.  
When we apply the dBB interpretation to the quantum cosmology, 
it naturally represents how the classical universe emerges 
for large cosmic scale factor\cite{Vink92,Horiguchi94,Alwis94,Kenmoku96}. 


In this paper, we apply the dBB interpretation to the quantum 
geometry of the 
Schwarzschild black hole, which is one of other interesting 
issues in the quantum gravity, because the quantum effect may become 
large near the singularity or the horizon. 
The quantum mechanical property of the black hole geometry may 
affects to the Hawking radiation\cite{Hawking75,Hajicek84,Tomimatsu92,Hosoya97} 
and to the phenomenon of the smearing of a black hole singularity
\cite{Hosoya95}. 
In order to treat the quantum theory of black hole,   
Nambu and Sasaki\cite{Nambu88} have proposed the canonical 
quantum gravity inside the horizon of the black hole and 
introduced the mass scale using a dust collapse model. 
Nakamura, Konno, Oshiro and Tomimatsu (NKOT) 
\cite{Nakamura93} solve the WD equation 
with the limitation of the mass eigenstate to study the 
quantum fluctuation of the horizon and found an    
exact wave function for the minisuperspace 
model of the interior geometry of the black hole. 
They researched the quantum effect on the horizon geometry 
on the basis of the WKB approach. 
The quantum fluctuation may become very large near the 
horizon, where the WKB method cannot be adopted. 
In order to estimate the quantum fluctuation quantitatively 
beyond the semi-classical region, we study the dBB interpretation 
for the wave function of the Schwarzschild black hole.          


In the classical theory, a spherically symmetric gravitational field in empty space 
must be static and its metric is given by the Schwarzschild solution 
by the Birkhoff theorem\cite{Weinberg}. 
In order to study the quantum theory, 
the canonical formalism must be formulated within 
the horizon region and the mass function is expressed by canonical variables.   
The mass eigenvalue equation plays an important role 
in order to argue the quantum theory of the empty space.
Instead of imposing the momentum constraint, 
the mass eigenvalue equation guarantees the diffeomorphism invariance 
not only inside but also the out side of the horizon.
The mass eigenvalue equation restricts the solution space of the WD equation 
strongly and plays the role of a kind of the initial or boundary 
condition.     
We use the canonical definition of the mass function considered 
by Fischler, Morgan, and Polchinski\cite{Fischler90} and 
by Kucha$\check{{\rm r}}$\cite{Kuchar94}, 
who has examined the canonical quantization formalism by Dirac\cite{Dirac64}, 
and by Arnowitt, Deser and Misner\cite{ADM62} for the spherically 
symmetric spacetime. They have shown that the mass function can be 
described by the canonical variables and is a constant of motion.
We solved the WD equation and the mass eigenstate equation simultaneously 
and obtained the general wave function, which is essentially same as 
that by NKOT but is taken account of the general form of operator ordering. 
We apply the dBB interpretation to this wave function and obtain the dBB 
trajectories on the minisuperspace, which represents the quantum feature 
of the geometry of the Schwarzschild black hole. We investigate the 
light ray on this quantum geometry. 


This paper is organized as follows.
In section 2, we review the canonical formalism in spherically 
symmetric spacetime. 
In section 3, we perform the canonical quantization of this system 
and obtain the wave function for the Schwarzschild black 
hole under a general consideration of the operator ordering. 
In section 4, we consider the implication of the wave function 
with the help of the dBB interpretation. Summary and discussion  
are given in section 5.      

%
\newsection{Canonical formalism in spherically symmetric spacetime }
%
In this section, we review the canonical formalism in the spherically 
symmetric spacetime in four dimension. 
The metric of the general spherically symmetric spacetime is 
\begin{equation}
 ds^2 = -u(r) dt^2 +  \f{\alpha(r)^2}{u(r)} dr^2  + r^2 d\Omega^2,
\end{equation}
where the each metric depends only on the space variable r and
$d\Omega^2 = d\theta^2 + \sin^2 \theta d\phi^2$ denotes the line element on the 
unit sphere. 
The unique solution of the vacuum Einstein equation for this metric is 
known as the Schwarzschild solution for  
\begin{eqnarray}
 \alpha = 1, \hspace{0.5cm} {\rm and}  \hspace{0.5cm}   u(r) = 1-\f{2Gm}{r},    
\end{eqnarray}
where $G$ denotes the gravitational constant.  
The integration constant $m$ represents the asymptotically observed mass   
of a spherically symmetric matter.  
The metric becomes singular when $r=0$ and $r=r_g \equiv 2Gm$.  
Here $r=0$ is a real physical singularity, whereas $r=r_g$ is a coordinate 
singularity and corresponds to the event horizon.   
Inside the black hole $(r < r_g )$, the metric $g_{tt}$ becomes   
positive and $g_{rr}$ negative, and  
so the role of time and space coordinate are interchanged. We denote this 
situation as 
\begin{equation}
 t \rightarrow R, \hspace{0.5cm}  
 r \rightarrow T, \hspace{0.5cm}   
 u(r) \rightarrow -U(T), \hspace{0.5cm}  
 \alpha(r) \rightarrow \alpha(T) . 
\end{equation} 
Then the interior metric of a spherically symmetric black hole is 
represented as
\begin{equation}
 ds^2 = -\f{\a(T) ^2}{U(T)}dT^2 + U(T) dR^2 + V(T) d\Omega^2,
\label{UV}
\end{equation}
where the range of the variables are $T>0,~-\infty < R < \infty$. 
The metric $V(T)$ is introduced in order to represents  
the quantum geometry of the spherically symmetric spacetime.
In this metric, the classical solution is written as: 
\begin{eqnarray}
 \a = 1 , 
\hspace{0.5cm}
  U= -(1-\f{r_g}{T}),  
\hspace{0.5cm}
  V^{1/2} = T.
\label{cl}
\end{eqnarray}

The vacuum Einstein action 
\begin{equation}
 S = \f{1}{16\pi G}\int  d^4x \sqrt{-^{(4)} g} \ ^{(4)}R
\label{Gravaction}
\end{equation}
is written in the form of (3+1) decomposition in terms of the ADM analysis 
as  
\begin{equation}
 S = \int dT L,
\end{equation}
where $L$ is the Lagrangian 
\begin{equation}
 L = \f{1}{16 \pi G} \int d^3 x N \sqrt{\ ^{(3)}g} 
    \bl(  K_{ab}K^{ab}  - (K^a_{~a})^2 + \ ^{(3)} R \br), 
\label{Lag(3+1)}
\end{equation}
where $N$ and $K_{ab}$ denote the lapse function and the extrinsic curvature, 
and 
$^{(3)} g$, 
and 
$^{(3)} R $ 
denote the metric and the curvature 
in three dimensional manifold respectively.
By inserting the metric in Eq.(\ref{UV}), 
the Lagrangian Eq.(\ref{Lag(3+1)}) becomes 
\begin{equation}
 L = \f{v_0}{4G} \bl[-\f{1}{\a}(\dot{U}\dot{V} + \f{U\dot{V}^2}{2V}) + 2\a \br].
\label{LagUV}
\end{equation}
where dot denotes the derivative with respect to $T$.  
The volume of the system $v_0 \equiv \int^{\infty}_{0} dR$ 
is treated to be finite.

Here we change the variables from $U,V$ to $z_+,z_-$ as follows:
\begin{eqnarray}
 z_+ \equiv U \sqrt{V}, \hspace{0.5cm}  
 z_- \equiv   \sqrt{V}.  
\label{z}
\end{eqnarray}
By using these new variables, the Lagrangian (Eq.(\ref{LagUV})) becomes 
simpler and symmetric form\cite{Suzuki96}
\begin{equation}
 L = \f{v_0}{2G} \bl[ -\f{1}{\a} \dot{z}_+ \dot{z}_- + \a \br].
\end{equation}
The canonical momentum conjugate to $z_+,\ z_-$ and 
$\a$ are obtained from this Lagragian:
\begin{eqnarray}
 {\it\Pi}_+ &\equiv& \f{\p L}{\p \dot{z}_+} = -\f{v_0}{2G\a} \dot{z}_-, 
\label{P+cl}\\
 {\it\Pi}_- &\equiv& \f{\p L}{\p \dot{z}_-} = -\f{v_0}{2G\a} \dot{z}_+, 
\label{P-cl}\\
 {\it\Pi}_{\a} &\equiv& \f{\p L}{\p \dot{\a}}  = 0.
\label{primary}
\end{eqnarray}
The variable $\a$ plays the role of the lapse function  
and so its canonical conjugate momentum (\ref{primary}) becomes zero,  
which is the primary constraint.  
The Hamiltonian for this system becomes of the form:
\begin{equation}
 H = -\a \bl[\f{2G}{v_0}{\it\Pi}_+ {\it\Pi}_- + \f{v_0}{2G} \br].
\label{clH}
\end{equation}
The development of the primary constraint (\ref{primary})  
by this Hamiltonian (\ref{clH}) yields the secondary constraint:
\bea
H|_{\a =1} \approx 0 ,
\label{Hconstraint}
\eea 
which is called the Hamiltonian constraint.

\vspace{0.5cm}
Next we discuss the mass of the system.
In the usual classical theory, the unique solution of 
the vacuum Einstein equation for the spherically symmetric spacetime 
is the Schwarzschild solution with one integration constant, which 
represents the mass of the black hole. 
In the canonical formalism for the spherically symmetric spacetime, 
the mass can be dynamical function, 
which has 
been introduced by Fischler, Morgan, and Polchinski\cite{Fischler90} and 
 extensively discussed by Kucha$\check{{\rm r}}$\cite{Kuchar94}. 
The spherically symmetric hypersurface on which the canonical data is given 
is supposed to be embedded in  a Schwarzschild black hole spacetime 
whose metrics are given by Eq.$(\ref{cl})$. 
This identification of the spacetime with the canonical 
data enables us to connect the Schwarzschild mass $M$ with the canonical data
on any small piece of a space-like hypersurface.
As results, the mass function is 
expressed by the canonical variables $z_+,z_-$ as   
\begin{equation}
 M=\f{2G}{v_0^2}~ z_+ ~({\it\Pi}_+)^2 + \f{z_-}{2G}.
\label{clM}
\end{equation}
The Poisson bracket of this mass function with  $H$ vanishes weakly:  
\begin{equation}
 \{ H,M \}_{\bf{p}} = -i\hbar \f{2G}{v_0^2} ~{\it\Pi}_+ ~H \approx 0, 
\end{equation}
which shows that the mass is a constant of motion.
Therefore we can consistently impose two equations: the Hamiltonian constraint 
and the mass constraint $M=m \ ({\rm constant})$ on canonical data.
\newsection{Quantization}
In this section, we proceed the canonical quantization treatment 
of the spherically symmetric spacetime, that is, the 
Schwarzschild black hole system.
We do not discuss the Hilbert space of the quantum gravity, and instead of this,  
we consider a general form of the operator ordering for 
the canonical operators.
 
In the Schr$\ddot{{\rm o}}$dinger representation, the canonical momenta are 
quantized as      
\begin{eqnarray}
 \hat{{\it\Pi}}_{+p} &\equiv& z^p_+ \hat{{\it\Pi}}_+ z^{-p}_+ = \hat{{\it\Pi}}_+ 
     + \f{i\hbar p}{z_+} \\
 \hat{{\it\Pi}}_{-s} &\equiv& (z_- - r_g)^s \hat{{\it\Pi}}_- (z_- - r_g)^{-s} 
     = \hat{{\it\Pi}}_- + \f{i\hbar s}{z_- - r_g},
\label{Pi+-}
\end{eqnarray}
where  
$\hat{{\it\Pi}}_+ $ and $\hat{{\it\Pi}}_- $ are usual differential operators
\bea
\hat{{\it\Pi}}_+ =-i\hbar \f{\p}{\p z_+}, \hspace{0.5cm}
\hat{{\it\Pi}}_- =-i\hbar \f{\p}{\p z_-}, 
\eea 
and $p$, $s$ are integers in order to take account of the operator ordering.  
The Hamiltonian 
constraint (\ref{Hconstraint}) 
gives a condition imposed on the state vector $\Psi$ 
\bea
\hat{H} \Psi 
= - \bl( \f{2G}{v_0} \hat{{\it\Pi}}_{+r}\hat{{\it\Pi}}_{-s} + \f{v_0}{2G} \br) \Psi 
=0,
\label{WD} 
\eea
where $r$ and $s$ are arbitrary integers. This is the WD equation for the geometry of the 
Schwarzschild spacetime. In addition, 
we confine the state vector to the mass eigenvalue equation. Therefore,     
\begin{eqnarray}
 (\hat{M}-m) \Psi = 
(\f{2G}{v_0^2}  \hat{{\it\Pi}}_{+p}z_+ \hat{{\it\Pi}}_{+q} + \f{z_-}{2G}  - m) \Psi 
=0,
\label{M}
\end{eqnarray}
where $p,q$ are arbitrary integers. 

 We note that the commutation relation between the Hamiltonian 
and the mass operator is calculated to be    
\begin{equation}
 [ \hat{H},~ \hat{M}-m ] = \f{-2i\hbar G}{v_0^2} \hat{{\it\Pi}}_{+r} \hat{H}
     + \f{4iG^2 \hbar^2}{v_0^3}(r-p)(r-q) \ \f{1}{ z^{2}_+}\  \hat{{\it\Pi}}_{-s}.
\end{equation}
For $r=p$ or $r=q$,  
the commutation relation vanishes weakly 
and the simultaneous requirement of 
the WD equation and the mass eigenvalue equation becomes compatible.  
In the following, we take the case of $r=q$. 
The case of $r=p$
is obtained by replacing $q \leftrightarrow p$ in the following calculation.
 
Instead of solving the mass eigenvalue equation directly, 
we consider the eigenvalue equation derived from the linear combination of the 
Hamiltonian and the mass operator: 
\bea
\bl(\hat{L} -\hbar \ (p-s)\br)\Psi =0 , 
\label{L}
\eea
where the operator $\hat{L} $ is defined as 
\bea
\hat{L}
&\equiv&  -2iG \bl(\f{1}{v_0} \hat{{\it\Pi}}_{+p} z_+ \hat{H} +\hat{{\it\Pi}}_{-s}(\hat{M}-m)\br) 
           +\hbar\ (p-s),
            \nonumber \\ 
&=&       i\bl( z_+ \hat{{\it\Pi}}_{+} -(z_- -r_g) \hat{{\it\Pi}}_{-} \br).
\label{defL}
\eea
As this equation is the first order differential equation, we can treat it easier.  
If we define the wave function $\psi$:
\begin{equation}
 \Psi \equiv z^q_+ (z_--r_g)^s \psi(z_+,z_-),
\label{psi}
\end{equation}
the equations we have to solve become the following two:
\bea
 \bl(\f{2G}{v_0}\hat{{\it\Pi}}_+\hat{{\it\Pi}}_-  + \f{v_0}{2G} \br) \psi 
= 0 
\label{h} ,
\eea
\bea
\bl(\hat{L} -\hbar\ (p-q) \br) \psi 
= 0.
\label{l}
\eea
If we make change of variables
\bea
 y \equiv \f{v_0}{G\hbar} \sqrt{-z_+/(z_- - r_g)},  \hspace{0.5cm}
 z \equiv  \f{v_0}{G\hbar}\sqrt{-z_+(z_- - r_g)},
\label{yz}
\eea
Eq.(\ref{l}) becomes 
\bea
 \bl[ y \f{\p}{\p y} - (p-q) \br] \psi (y,z) = 0,
\eea
whose  general solution is given by  
\bea
 \psi(y,z) =  y^{p-q} u(z),
\label{u}
\eea
where $u(z)$ is an arbitraly function of $z$.  
By substituting  the new variables (\ref{yz}) and 
the wave function (\ref{u}) into Eq.(\ref{h}),
the WD equation is reduced to 
\bea
  \bl[\f{1}{z}\f{\p}{\p z} z \f{\p}{\p z}
   - \f{y^2}{z^2} \bl(\f{1}{y}\f{\p}{\p y} y \f{\p}{\p y}\br) 
   +1 \br] y^{p-q} u(z)  
 &=& y^{p-q}\bl[\f{1}{z}\f{\p}{\p z}z \f{\p}{\p z} 
   - \f{(p-q)^2}{z^2}  + 1 \br] u(z)
\non \\
 &=& 0.  
\label{Bessel}
\eea
The equation for $u$ is the Bessel's differential equation with order $p-q$.
Then we obtain the eigen function  for the WD and the mass eigenvalue equations  
\bea
\Psi = y^{p-s}z^{q+s} \bl[c_1 H^{(1)}_{p-q}(z) + c_2 H^{(2)}_{p-q}(z)\br] ,
\label{solution}
\eea
where $c_1,~c_2$ are integration constants.
The Hankel functions $H^{(1)},~H^{(2)}$ are 
linearly independent and complex conjugate each other.

NKOT have derived the quantum wave function for the Schwarzschild 
black hole.
As the variables $y,~z$ in Eq.(\ref{yz}) take real values for the physical  
metrics $U,~V$ in Eq.(\ref{UV}),  
the complex phase factor comes only from the Hankel function. 
Note that our solution includes NKOT's solution\cite{NKOTsolution}.    

%
\newsection{Implication of wave function}
%
NKOT have discussed the quantum evolution of the metric variables 
from the wave function of the black hole in the WKB approach\cite{Nakamura93}.   
They further argued the possibility of a tunneling solution across the horizon, 
where the quantum fluctuation becomes large and the WKB approach is not applicable. 
In order to make the quantitative estimation of the quantum fluctuation, 
we apply the de Broglie-Bohm (dBB) interpretation to 
the wave function of the quantum black hole which is obtained in the previous 
section.
This dBB interpretation has been applied to quantum cosmology by 
several authors
in order to solve 
the problems of time and the observer\cite{Alwis94,Vink92,Horiguchi94,Kenmoku96} . 

\vspace{0.5cm} 

In the following, we apply the dBB interpretation 
to the wave function of the spherically symmetric black hole.  
For this purpose, we rewrite the WD equation and the mass eigen equation explicitly  
in terms of the real phase and the real amplitude of the wave function:  
\bea  
\Psi(z_+,z_-)= R(z_+,z_-)\exp(iS(z_+,z_-)/\hbar). 
\label{RandS}
\eea
Inserting this expression into the WD equation (\ref{WD}), 
we obtain the real part equation:  
\bea
\f{2G}{v_0}~\f{\p S}{\p z_+} \f{\p S}{\p z_-} + \f{v_0}{2G} + Q  = 0, 
\label{reWD} 
\eea
where 
$Q$ denotes the quantum potential for the quantum black hole as 
\bea
Q = -\f{2G\hbar^2}{v_0 R}(\f{\p}{\p z_+}-\f{q}{z_+})
                        (\f{\p}{\p z_-}-\f{s}{z_- -r_g})R,  
\label{QQ}
\eea 
and the imaginary part equation:
\bea
(\f{\p}{\p z_+}-\f{q}{z_+})(R^2 \f{\p S}{\p z_-} ) +
(\f{\p}{\p z_-}-\f{s}{z_- -r_g})(R^2 \f{\p S}{\p z_+} )
=0.
\label{imWD}
\eea 
If $Q$ tends to zero, the real part of the WD equation, (\ref{reWD}) 
is reduced to the Hamilton-Jacobi equation for the black hole. 
This means that the quantum potential indicates the quantum effect quantitatively.  
In our quantum back hole model, we also demanded the mass eigenvalue 
equation (\ref{M}). 
The real part of this equation can be rewritten as
\bea
\f{2G}{v_0^2} ~{z_+}~ \bl( \f{\p S}{\p z_+} \br)^2 + \f{z_- -r_g}{2G} + M_Q = 0,
\label{reM} 
\eea
where $M_Q$ is defined as   
\bea
 M_Q = -\f{2G\hbar^2}{v_0^2 R} 
         (\f{\p}{\p z_+}-\f{p}{z_+}) (z_+
         (\f{\p}{\p z_+}-\f{q}{z_+}) R),
\label{MQ}
\eea
and the imaginary part is given as 
\bea
 \bl(\f{\p}{\p z_+} -\f{q+p}{z_+} \br)(R^2~ z_+ ~\f{\p S}{\p z_+}) = 0.
\label{imM}
\eea
If $M_Q$ tends to zero, Eq.(\ref{reM}) is reduced to the classical relation 
: $M$ in Eq.(\ref{clM}) equals to a constant $m$.  
Therefore $M_Q$ represents the quantum effect to the mass function. 
The operator $\hat{L}$ in Eq.(\ref{defL}), which is the linear combination of 
the Hamiltonian and the mass operator, is easy to analyze, 
because it is the first order differential operator. 
Its real and imaginary parts are rewritten respectively as 
\bea
\bl[z_+\f{\p}{\p z_+} - (z_- - r_g)\f{\p}{\p z_-}-p+s\br]R=0, \\
\label{reL}
\bl[z_+\f{\p}{\p z_+} - (z_- - r_g)\f{\p}{\p z_-}\br]S =0 .
\label{imL}
\eea
Using Eq.(\ref{reL}), 
the quantum effect of the mass operator $M_Q$ can be expressed to be proportional to the 
quantum potential $Q$:  
\bea
M_Q=\f{z_- - r_g}{v_0}Q,
\eea
so that all the quantum effect vanishes when $Q$ tends to zero.

Here we introduce the dBB interpretation, which is based on the following 
assumptions:

\noindent
(1) \ \ The trajectory picture on the minisuperspace is introduced.   
The momenta of the quantum geometry of the black holes are assumed to be given by    
\bea
  {\it\Pi}_+ &=& -\f{v_0}{2G}\dot{Z}_- = 
\br. \f{\p S}{\p z_+}\bl|_{z_+ = Z_+, z_- = Z_-} 
\label{P+} , \\
  {\it\Pi}_- &=& -\f{v_0}{2G}\dot{Z}_+ = 
\br. \f{\p S}{\p z_-}\bl|_{z_+ = Z_+, z_- = Z_-},
\label{P-}
\eea 
where the velocities are determined by the classical relations 
(\ref{P+cl}) and (\ref{P-cl}) with $\a=1$. 
The trajectories $Z_+,~Z_-$ are obtained by integrating 
Eqs. (\ref{P+}), (\ref{P-}). 
We emphasize that they  are the differential equations with respect to 
$T$, which is the notion of time parameter appeared in Eq.(\ref{UV}),  
so that the notion of time is introduced even if the 
WD equation does not depend on the time explicitly. 
We shall call these trajectories as the dBB trajectories hereafter. 

\noindent
(2) \ \  Consider a statistical ensemble of the trajectories. 
The probability distribution is assumed to be given by $R^2$. Note 
that the imaginary part equation guarantees the continuity for the probability 
density $R^2$. 

\noindent
(3) \ \ The quantum potential $Q$ denotes the quantum effect 
quantitatively. This is not a usual local potential because 
it is defined through a wave function. 
If $Q$ is negligible compared with the classical 
potential or kinetic term, the trajectory coincides with the classical one.
This means that the quantum system behaves like classical one spontaneously.  
It is indeed this situation what we call "classical". 
Therefore, we need no classical observer which is an essential element  
in Copenhagen interpretation.  
On the other hand, if $Q$ cannot be negligible, the dBB trajectory 
is modified by any quantum probe. 
In this case, we should explicitly include a specific quantum observer    
in the dynamical system to get information of the dBB trajectory.   
When the observer become classical, it plays the role of an observer in 
Copenhagen interpretation which is classical by nature.

We note that in dBB interpretation, an existance in the nature is a dBB 
trajectory, and therefore, predictability of quantum 
mechanics is based on probability distribution of their statistical ensemble.  
As a result, the reduction of 
the wave function or the loss of the quantum coherence 
by the measurement is not the matter with the dBB interpretation.

\vspace{0.5cm}
Next we insert the explicit form of the Hankel function (\ref{solution})  
 into the general form of the wave function (\ref{RandS}):  
\bea   
\Psi=y^{p-s}z^{q+s}H^{(2)}_{\nu}(z).  
\label{hankel} 
\eea 
Here we have chosen the Hankel function of the second kind 
$H^{(2)}$
 with the index $\nu \equiv p-q$, which denotes the 
freedom of the operator ordering,    
since from the dBB point of view it corresponds to the classical relation 
$V^{1/2}=T$ in Eq.(\ref{cl}) in semi-classical region, as will be seen below (Fig.2).  
Physically the selection of $H^{(2)}$
corresponds to solve the trajectory 
from the singularity at the origin to  
outside.   
On the ansatz the dBB trajectory for the wave function of 
the linear combination of both $H^{(1)}$
and $H^{(2)}$ in Eq.(\ref{solution}) is shown not to 
approach a classical trajectory.

After inserting Eq. (\ref{hankel}),  
the quantum potential is expressed by the Hankel function as  
\bea
 Q = - \bl. \f{v_0}{2G} \bl( 1 - \f{4}{\pi^2 z^2 
               | H^{(2)}_\nu (z)|^4} \br) . 
\label{QH}
\eea
Similarly, the continuity equations (\ref{imWD}), (\ref{imM}) 
and (\ref{imL}) are reduced to one equation as 
\bea
 \f{\p}{\p z} (~z | H^{(2)}_\nu (z) |^2 ~\f{\p}{\p z} S ) = 0,
\eea 
which corresponds to the identity of the Hankel function: 
\bea
H^{(2)} \f{\p}{\p z} H^{(1)} -H^{(1)} \f{\p}{\p z}  H^{(2)} 
= \f{4i}{\pi z}.
\label{identity}
\eea 
Noticing that the phase $S$ comes only from the Hankel function 
and that the derivative of the phase can be expressed by the identity  
equation (\ref{identity}), 
a couple of equations on the velocities (\ref{P+}) and (\ref{P-}) are 
obtained as
\bea
\dot{Z}_- &=&\f{2\hbar G}{\pi v_0}~\f{1}{Z_+ | H^{(2)}_\nu (Z)|^2} , 
  \label{Z-} \\
\dot{Z}_+ &=&\f{2\hbar G}{\pi v_0}~\f{1}{(Z_- -r_g) | H^{(2)}_\nu (Z)|^2}.
  \label{Z+}
\eea
We take the ratio of Eq.(\ref{Z-}) to Eq.(\ref{Z+}) 
to cancel the time dependence, and integrate it to 
get the $Z_+-Z_-$ relation:
\begin{equation}
 Z_+ = c_0 (Z_- - r_g),
\label{+-rel}
\end{equation}
where $c_0$ is the integration constant.
With the choice of  
$c_0=-1$ this relation is translated back to that of the original 
metric variables $U,V$ in Eq.(\ref{UV}):
\begin{equation}
 U = -(1-\f{r_g}{V^{1/2}}),  
\label{UVrel}
\end{equation}
which corresponds to the classical relation in Eq.(\ref{cl}).  
The $U-V^{1/2}$ relation is shown in Fig.1, where the boundary of the classical region 
is also shown. 
\begin{center}
===========\\
{\rm Fig.1}\\
===========
\end{center} 
We take the natural units $c=\hbar =G=1$ and $v_0=2$ in 
this and the following figures. 
  
Using the $U-V^{1/2}$ relation (\ref{Z-}), the remaining independent $T-V^{1/2}$
 relation is obtained 
in the integral form:
\bea
T=\f{\pi}{2}\int Z | H^{(2)}_\nu (Z)|^2 dV^{1/2}
\ \ \ {\rm with}\ \ \ Z=\f{v_0}{G\hbar}|V^{1/2}-r_g|.  
\label{TV}
\eea
Considering the asymptotic form of the Hankel function 
\bea
H^{(2)}_\nu (Z) \rightarrow \exp (-iZ)~\sqrt{\f{2}{\pi Z}} \hspace{0.5cm} 
{\rm for} \hspace{0.5cm} Z>>1,
\label{asym}
\eea
we can show that the $T-V^{1/2}$ relation approaches the classical relation 
$T=V^{1/2}$ in semi-classical region.
Near the horizon $Z\approx 0$, 
the Hankel function is approximately expressed as 
\bea
 | H^{(2)}_\nu (Z)|^2 \approx 
\left\{
\begin{array}{lr} 
\displaystyle \f{4}{\pi^2}~(\ln Z)^2                     
 \ \ \ \ \  {\rm for \ \  {\it \nu =0}} \\
\\
\displaystyle \f{2^{2\nu }{(\nu -1)!}^2}{\pi^2 Z^{2\nu}} 
 \ \ \ {\rm for \ \ {\it \nu}  = positive \ \ integer} .
\end{array}   
\right.
\label{approx}
\eea 
In the following, we discuss the case of $\nu=0$ and $\nu=1$.  
In the case for $\nu > 1$ the behavior of the physical quantities 
is similar to the case for $\nu=1$. 
We estimate the 
$T-V^{1/2}$
 relation (\ref{TV}) near the horizon using 
the approximate equation (\ref{approx}) as 
\bea
T-T_0 \approx 
\left\{
\begin{array}{lr}
\displaystyle - \f{v_0}{\pi G\hbar}~(V^{1/2}-r_g)^2~(\ln{|V^{1/2}-r_g|})^2       
& \ \ \ {\rm for}\ \ \ \nu=0 \\
\\ 
\displaystyle - \f{2G\hbar}{ \pi v_0}~\ln{|V^{1/2}-r_g|}                      
& \ \ \ {\rm for}\ \ \ \nu=1 ,
\end{array}   
\right.
\eea   
where $T_0$ is the integration constant.
From these expressions, we can see that 
$T$ shows flat behavior for $\nu=0$ and steep behavior for $\nu =1$
near the horizon $V^{1/2}\approx r_g$.
In Figs.2(a) and (b), numerical estimations on the $T-V^{1/2}$
 relation (\ref{TV}) are shown.
\begin{center}
=======================\\
{\rm Figs.2(a) and (b)}\\
=======================
\end{center} 
Fig.2 implies that the net region of the horizon  
extends widely for $\nu=1$ whereas some part of the horizon region 
is cut off for $\nu=0$, comparing with the classical 
relation $V^{1/2}=T$. We note that the inner 
metric for $\nu=1$ cannot connect 
the outside metric as $T$ diverges near the horizon. 
In Fig.2 (b), the double wavy mark denotes this discontinuity 
between the inside and the outside.  
The $U-T$ relation 
can be obtained from the $U-V^{1/2}$ relation in Fig.1 
combining $T-V^{1/2}$ relation in Fig.2. 

\vspace{0.5cm}
We next estimate the quantum potential $Q$, (\ref{QH}). 
Using the asymptotic behavior of the Hankel function in Eq.(\ref{asym}), 
the quantum potential becomes zero for $Z>>1$. The 
non-zero structure, which shows the quantum effects,  
exists only near the horizon $Z\approx 0$. The 
quantum potential, Eq.(\ref{QH}) is estimated approximately near the horizon, 
using the relation (\ref{approx}), as 
\bea
 Q \approx 
\left\{
\begin{array}{lr}
\displaystyle  \f{{\pi}^2 v_0}{8G} ~\f{1}{Z^2~(\ln{Z})^4}       
& \ \ \ {\rm for}\ \ \ \nu=0 \\ 
\\ 
\displaystyle \ \ \ - \f{v_0}{2G}                                   
& \ \ \ {\rm for}\ \ \ \nu = 1 . 
\end{array}   
\right.
\eea 
The quantum potential diverges positively for $\nu=0$ and 
is negative finite for $\nu = 1$. In Figs.3(a) and (b), 
the graphical representation of the quantum potential 
is shown.
\begin{center}
=======================\\
{\rm Figs.3(a) and (b)}\\
=======================
\end{center}
The quantum effect behaves very differently for each value of the index of the 
operator ordering $\nu=0$ and $\nu=1$.   

\vspace{0.5cm}

Now we consider the horizons. 
The event horizon, which has the global meaning, is located at  
 the null surface $U=0$. On the other hand, the apparent 
horizon, which has the local meaning, is defined by 
the expansion for outgoing null rays becomes zero:  
$\theta_+=0$. The product of the expansion for outgoing and incoming null 
rays can be written as\cite{Nakamura93,York83} 
\bea
\theta_- \theta_+ = UV^{-1}\left({\dot{\sqrt{V}}}\right)^2 
= \f{1}{V} (1-\f{2G\hat{M}}{V^{1/2}}). 
\label{theta+-}
\eea
In the classical theory, the apparent horizon takes on 
 ${V^{1/2}}=2Gm$ and agrees with the event horizon through 
the relation $U=-V^{1/2}+2Gm$. In quantum theory, $\theta_-\theta_+$ is a 
operator, and the quantum fluctuation becomes very large near 
$V^{1/2}=2Gm$. Then the estimation of the apparent horizon is not 
straightforward. Followed by NKOT\cite{Nakamura93}, we require 
the apparent horizon in the quantum theory as 
\bea
\theta_- \theta_+ \ \Psi = 0,
\label{apparent}
\eea
 where $\Psi$ denotes the simultaneous solution of the WD equation 
and the mass eigenstate equation. 
By this requirement the mass operator $\hat{M}$ in Eq.(\ref{theta+-}) is reduced to the 
eigen value $m$ and    
the quantum apparent horizon takes on the classical value $V^{1/2}=2Gm$. 
Then the relation $U-V^{1/2}$, Eq.(\ref{UVrel}) 
means that the two horizons, the event horizon and the apparent horizon, coincide 
from  the view point of the dBB interpretation. 

\vspace{0.5cm}
In order to understand the property of the horizon, 
we consider the motion of the light ray on the geometry of 
the quantum black holes. 
In principle, the light ray 
as well as the black hole geometry should be treated in the quantum theory. 
It is because the massless particle curves the spacetime where it propagates, 
while it propagates in the curved spasetime. 
Instead of this, we treat the light ray as a 
test particle in order to probe the quantum property of the geometry. 
It means that we consider the low energy limit of the light ray 
and the quantum geometry does not take the influence from 
this test light ray.  
 
The null condition for the metric in Eq.(\ref{UV}) becomes 
\begin{equation}
\f{dR}{dT}= \pm \f{1}{U}  \ \ \   {\rm with} \ \ \ U=\f{r_g}{V^{1/2}}-1.
\label{RT}
\end{equation}
We take the variable $V^{1/2}$ instead of $T$ as
a independent variable, where the $T-V^{1/2}$ relation is seen  
in Eq.(\ref{TV}).  Then we obtain the integral expression 
for the light ray as
\bea
R&=& \pm \int \f{1}{U}~\left( \f{dV^{1/2}}{dT} \right) ^{-1}~dV^{1/2} \non \\
 &=& \pm \f{\pi v_0}{2G\hbar} ~\int V^{1/2}~|H^{(2)}(Z)|^2 dV^{1/2}
\ \ \ {\rm with}\ \ \ Z=\f{v_0}{G\hbar}(V^{1/2}-r_g).
\label{RV}
\eea 
The approximate behavior 
near the horizon $V^{1/2} \simeq r_g $
 of this equation is estimated using Eq.(\ref{approx}) as 
\bea
R -R_0 \simeq
\left\{
\begin{array}{lr}
\displaystyle \mp \f{v_0}{\pi G\hbar}(V-r_g^2)\ (\ln{|V_{1/2}-r_g|)^2}
& \ \ \ {\rm for} \ \ \ \nu=0 \\ 
\\
\displaystyle \mp \f{G^2\hbar m}{4\pi v_0} ~\f{1}{V^{1/2}-r_g} 
     & \ \ \ {\rm for} \ \ \ \nu = 1 ,
\label{appRV}     
\end{array}   
\right.
\eea
where $R_0$ denote the integration constant. 
In Fig.4, the numerical estimations of $R-V^{1/2}$ relation  
for the light ray (\ref{RV}) are shown.  
\begin{center}
========================\\
{\rm Figs.4 (a) and (b)}\\
========================
\end{center}
For the comparison, the light ray in the classical case 
\bea
R_{\rm cl} = -T -2 \ln{|V^{1/2}-r_g|} + 2\ln{r_g} 
\hspace{0.3cm}{\rm and}\hspace{0.3cm}V^{1/2}=T ,   
\eea
is also shown in the figures. 
The integration constant is fixed so that the light ray on the dBB trajectory 
and that on the classical trajectory coincide at the origin $(T=0)$ and 
the infinity $(T=\infty)$.
The classical ray is modified near the horizon region. 
In case of $\nu=0$ the light ray forms a cusp of a finite height of $R$ at 
 the horizon position $V^{1/2}=2Gm$. 
In case of $\nu=1$ though the light ray behaves like  
to that in the classical case, 
it diverges as the inverse power $(V^{1/2}-r_g)^{-1}$ near the horizon 
in stead of logarithm in the classical case. 
In this case  the horizon region is just expanded. 

%
\newsection{Summary and Discussion}
%
We have studied the dBB interpretation for the quantum theory of 
the black hole geometry. 
We have estimated the dBB trajectories and the quantum potential. 
The dBB trajectories have been obtained from the phase and the quantum potential 
from the amplitude of the wave function. 
By considering both quantities, we can get the total picture of the quantum 
black holes. 


Here we compare the dBB interpretation with the WKB approach. 
The WKB approach covers the semi-classical region. 
However by means of the dBB interpretation we can define the dBB 
trajectories not only in the classical region but also in the quantum 
region and get the glabal picture through the dBB trajectories.   
The quantum effect can be estimated quantitatively and continuously 
from the quantum potential and from the difference 
of the dBB trajectories with the classical ones. 
In the case that the quantum potential is negligible, 
the trajectories in the dBB interpretation agrees with those in the WKB approach, but  
its interpretation is different. 
For example, 
wave functions are peaked about strong correlation between coordinates and momenta 
along the trajectories in the WKB approach discussed by Halliwell\cite{Halliwell87} 
and Kadama\cite{Kodama88}.   
On the other hand, the trajectories in the dBB interpretation 
have the causal meaning  
and therefore approach the classical ones without measurement process.   
 
\vspace{0.5cm}   
In the region where the quantum effects cannot be negligible, 
any probe which plays the role of the observation in Copenhagen interpretation 
may affect to the dBB trajectories. 
We have considered the motion of the light ray on the quantum geometry  
in order to study the quantum effect, which 
becomes very large near the horizon. In the evaluation in Figs. 4 (a) and (b), 
we treat the light ray as a test particle, which means 
that it does not affect to the geometry.
In principle, the light ray 
as well as the black hole geometry should treat in the quantum theory,  
because the light ray curves the spacetime which it propagates. 
In the appendix, a model for the quantum theory of the massless particle 
and the black hole geometry is considered and its connection to the 
calculation of the light ray in section 4 is studied.


We concluded that  the classical $U-V^{1/2}$ relation (\ref{UVrel}) 
holds also in the quantum case on the dBB trajectory. 
Then the event horizon and the apparent horizon  
which is defined in Eq.(\ref{apparent}) are coincide.
The reason why the classical and the quantum $U-V^{1/2}$ relation 
become same is that two constraints,  
the Hamiltonian constraint and the mass constraint,  
are imposed on the wave function  
and then   
 the phase factor $S$ of the wave function 
becomes the function of the only one variable 
$z$ in Eq.(\ref{yz}) and so the ratio of 
the two equations defining the momenta of the dBB trajectories,  
(\ref{P+}) and (\ref{P-}) 
 can be calculated unambiguously to derive the $U-V^{1/2}$ relation.
The possibility of the inequality of the event horizon to  
the apparent horizon was discussed by NKOT\cite{Nakamura93}. 
Their main interest lays in the tunneling effect from the outside 
to the inside of the horizon. 
They argued that if the apparent horizon does not 
coincide with the event horizon, 
 the $U-V^{1/2}$
 curve outside the horizon 
 can be connected to 
that inside the horizon 
through the tunneling of the forbidden region.

In our analysis, the quantum fluctuation largely depends upon the ordering parameter. 
In case of the ordering parameter $\nu\geq 1$, 
the net horizon region is enlarged to infinity as seen in Fig.2 (b) 
and the causal connection between the inside and the outside 
of the horizon is cut off. 
Its mathematical reason is the strong singular behavior of the 
Hankel function at the $z=0$. 
On the other hand, in case of $\nu=0$, 
the light ray on the 
quantum geometry of the black hole forms a cups and reaches the horizon region 
within the finite $R$,
 which plays the role of time originally. 
So the $\nu=0$ case is specially interesting. 
We note also that if the Hermiticity is required to the 
quantum operators such as the Hamiltonian and the mass operator, 
 even though the Hilbert space of quantum gravity is not 
well understood yet,  
the relation among the ordering indexes $\nu =p-q=0$
 is preferred and 
then the quantum solution (\ref{solution}) is expressed by 
the Hankel function with the order 
$\nu=0$.

\vs{0.3}
Concerning to our work, an interesting phenomenon is on 
the Hawking radiation\cite{Hawking75}. 
As we have obtained the quantum geometry of the black hole, 
it will give the effects on the scattering or the radiation of the 
electromagnetic and matter fields. 
Recently using the Vaidya metric 
black hole radiation in quantum geometry 
was discussed by Tomimatsu\cite{Tomimatsu92} and Hosoya and Oda\cite{Hosoya97}.
The black hole radiation in our approach remains future problem. 
    
\newpage 
\appendix
\section{Model for massless particle }
We consider a quantum model of the massless particle and the black hole 
geometry. The action consists of two parts: the gravity part  
(\ref{Gravaction}) and the massless particle action $S_M$, which is
\bea
S_M= \displaystyle \int d^4x \sqrt{-^{(4)}g} \ \f{1}{8\pi}
\ ^{(4)}g^{\mu \nu}
\ ^{(4)}g_{\sigma \rho}
\            (\p_{\mu} X^{\sigma})(\p_{\nu} X^{\rho}),
\label{Maction}
\eea 
where $X$ denotes the coordinate of the massless particle,  
which are assumed to take only the time and the space components as 
 functions of $T$, 
\bea
X^{\sigma} = (X^0(T)/\alpha, X^1(T),0,0) .
\eea
We substitute insert the metrics Eq.(\ref{UV}) into the matter action (\ref{Maction}) 
 and change the variables to symmetric ones $z_+,z_-$, Eq.(\ref{z}). 
Then the matter action becomes
\bea
S_M = \displaystyle \int dt dr \f{V}{2\alpha}((\dot{X^0})^2 
                                             -{U^2}(\dot{X^1})^2).
\eea 
 The canonical momenta conjugate to $X^0$ and $X^1$ are calculated as 
\bea
P_0 = v_0 \f{V}{\alpha} \dot{X^0} , \hspace{0.5cm}
P_1 = -v_0 \f{V U^2}{\alpha} \dot{X^1} .
\label{P0P1} 
\eea
The massless particle Hamiltonian $H_M$ is  obtained as
\bea
H_M = \f{\alpha}{2v_0}\ (\f{1}{z_-^2}\ P_0^2 - \f{1}{z_+^2}\  P_1^2).
\eea 
Again the total Hamiltonian constraint is obtained 
\bea
H + H_M \approx 0 \hspace{0.3cm} {\rm with} \hspace{0.3cm} \alpha = 1.
\eea
After the canonical quantization
 $P_0=-i\hbar \p_0$ and $P_1=-i\hbar \p_1$, 
we obtain the WD equation for the wave function 
$\Psi(z_+,z_-,X^0,X^1)$ as
\bea
(\hat{H} + \hat{H_M}) \ \Psi(z_+,z_-,X^0,X^1) = 0.      
\eea
By solving this WD equation we get the total quantum picture of the massless 
particle and the black hole geometry. 
Now we investigate the relation of this equation to the analysis which 
has been done in the section 4. 
First we make the approximation that the wave function is assumed to be of the 
form in the separation of the variables 
\bea
\Psi(z_+, z_-, X^0, X^1) \simeq \Psi(z_+, z_-)\  \Phi(X^0,X^1). 
\eea 
In this approximation, the wave functions $\Psi(z_+, z_-)$ and 
$\Phi(X^0,X^1)$ satisfy the WD equation separately, which are 
Eq.(\ref{WD}) and 
\bea
-\f{\hbar^2}{2 v_0 z_-}\ (\p_0^2 - \f{z_-^2}{z_+^2}\ \p_1^2)\  \Phi(X^0,X^1) =0,
\eea 
where $z_+$ and $z_-$ are assumed to take values of the dBB trajectories:  
(\ref{P+}) and (\ref{P-}).
We further make the WKB approximation for the phase of the wave function 
$\displaystyle \Phi(X^0,X^1) \simeq \exp{\f{i S_M}{\hbar}}$ and 
identify its derivative with the momenta (\ref{P0P1}), and 
get the equation for the massless particle as 
\bea
\f{d X^1}{dX^0} = \f{z_-}{z_+} = \f{1}{U}. 
\eea
Identifying $X^0\equiv \lambda T$ and $X^1\equiv \lambda R$, 
where $\lambda$ is introduced as a scaling parameter,  we obtain 
the light ray equation (\ref{RT}) in section 4.
In this approximation, the massless particle and the geometry 
do not give the influence each other. 
If we improve the approximation to include the correlation, 
the mass operator cannot take constant value. 
The change of the mass operator is expressed through the commutation relation 
between the mass operator and the Hamiltonian as 
\bea
\dot{\hat{M}} &=&
 \f{1}{i\hbar}[\hat{H} + \Delta \hat{H} , \ \hat{M}] \nonumber \\
&=&    \f{-2G}{v_0^2} \hat{{\it\Pi}}_{+} \hat{H}
     + \f{2G}{v_0^2}({\it \Pi}_+ H_M + \f{1}{v_0} \ 
       (z_+^{-2}{\it\Pi}_+ {\it\Pi}_+ z_+^{-2}) P_1^2), \nonumber \\ 
 &\approx& 
 \f{2G}{v_0^2}({\it \Pi}_+ H_M + \f{1}{v_0} \ 
       (z_+^{-2}{\it\Pi}_+ {\it\Pi}_+ z_+^{-2}) P_1^2),  
\eea
where we take the Weyl operator ordering. 
This equation shows that the mass function changes according to the interaction 
of the masslass particle and the geometry.   
The analysis of this model remains our future problem. 

\clearpage

\clearpage
\begin{center} 
{\large{Figure Captions}}
\end{center}
\begin{description}
\item[Fig.1] \ \ 
 The $U-V^{1/2}$ relation is shown.
The semi-classical region is bounded by dashed line. 
The event horizon $(U=0)$ and the apparent horizon 
$(V^{1/2}=2Gm)$ coincide on the dBB trajectory. 
The natural units $c=\hbar=G=1$  and $v_0=2$ are taken.\\
\item[Fig.2] \ \ 
 The $T-V^{1/2}$ relation is shown. 
The ordering parameter is taken $\nu=0$ in (a) 
and $\nu=1$ in (b). 
The classical relation 
$T=V^{1/2}$ is denoted by 
dashed line. 
The double wavy mark in (b) indicates the discontinuity between 
the inside and the outside of the horizon. 
Axis unit is same as in Fig.1. \\
\item[Fig.3] \ \ 
The quantum potential $Q$ is shown with respect to the variable $V^{1/2}$. 
The ordering parameter is taken $\nu=0$ in (a) 
and $\nu=1$ in (b). 
The large positive effect is appeared for $\nu=0$ and 
the negative constant effect is appeared for $\nu=1$  
near the horizon.
Axis unit is same as in Fig.1. \\ 
\item[Fig.4] \ \ 
The light ray on the dBB trajectory is shown. 
The light ray on the classical geometry is indicated by dashed line. 
The ordering parameter is taken $\nu=0$ in (a) and 
$\nu=1$ in (b). The light ray forms a cups and reaches to the horizon at 
finite $R$ in case of $\nu=0$. 
The light ray in case of $\nu=1$ behaves like that in the classical case. 
Axis unit is same as in Fig.1. 
\end{description}  

\pagebreak
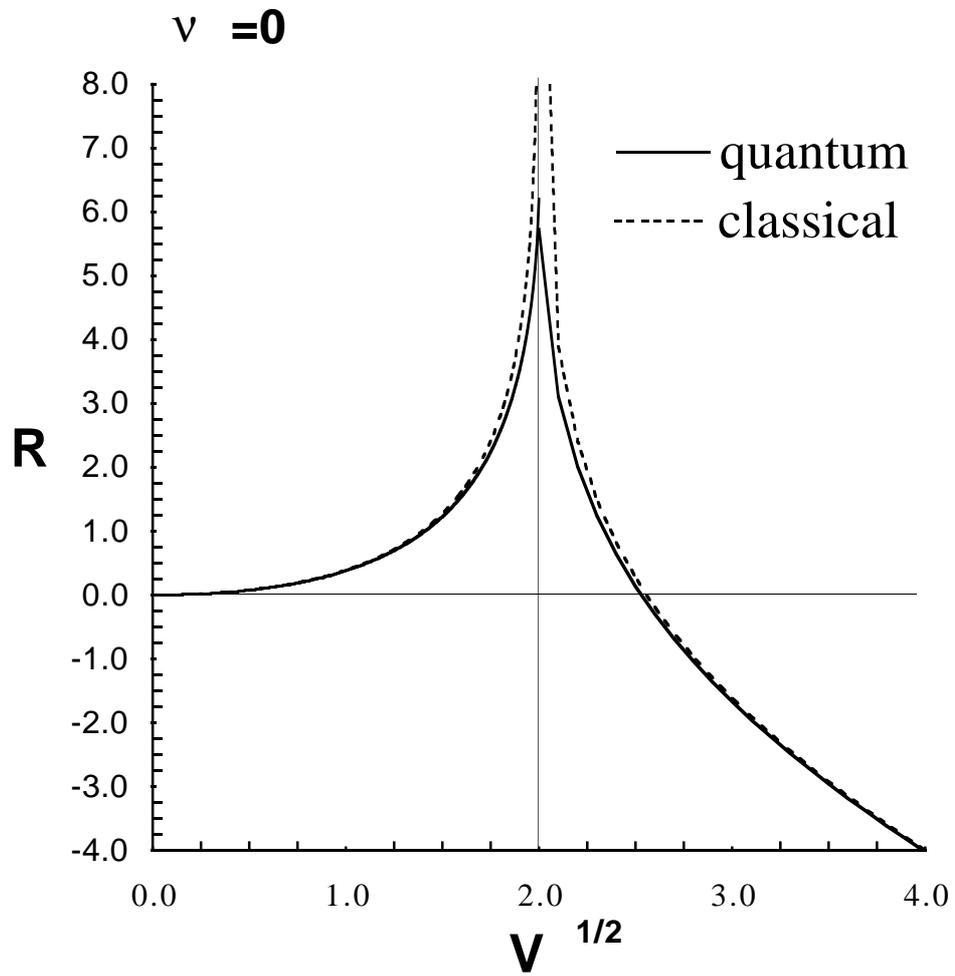
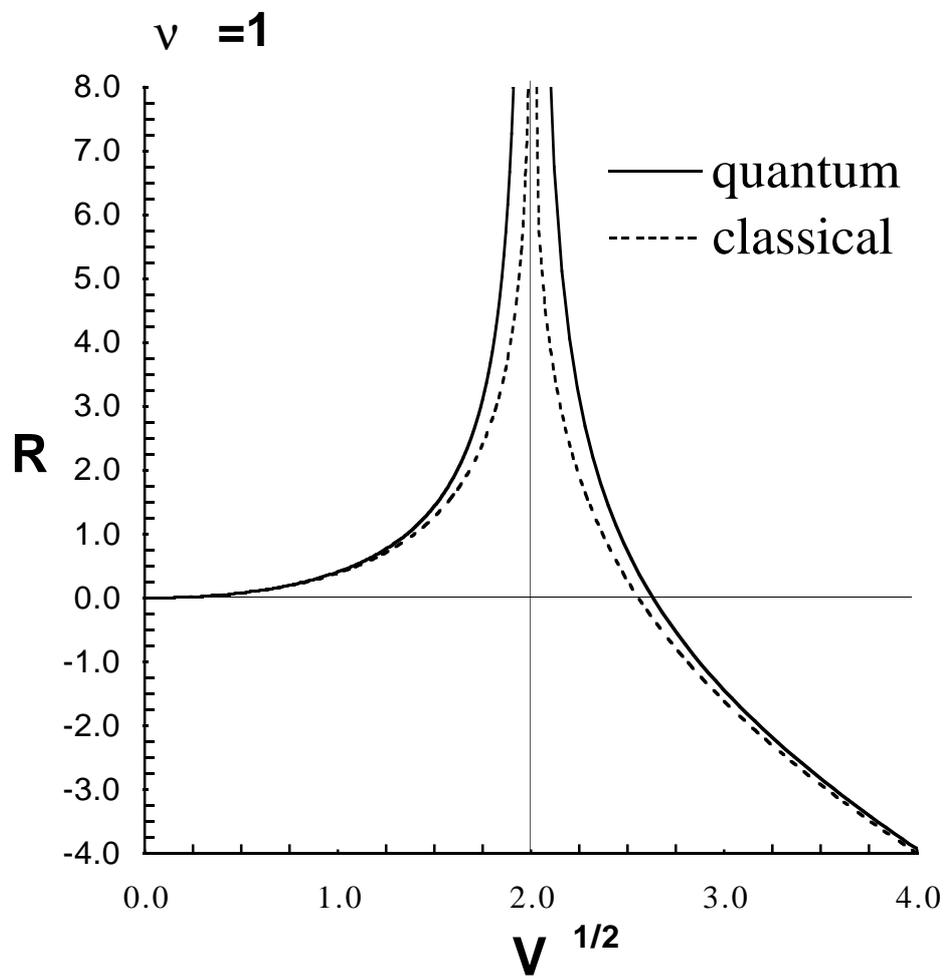
\end{document}